\documentclass[twocolumn]{aastex63}
\usepackage{graphicx}
\usepackage{amsmath}
\usepackage{amssymb}
\usepackage{hyperref}

\def\gtrsim{\mathrel{\hbox{\rlap{\hbox{\lower4pt\hbox{$\sim$}}}\hbox{$>$}}}}
\def\lesssim{\mathrel{\hbox{\rlap{\hbox{\lower4pt\hbox{$\sim$}}}\hbox{$<$}}}}
\def\gtrsim{\mathrel{\hbox{\rlap{\hbox{\lower4pt\hbox{$\sim$}}}\hbox{$>$}}}}
\def\farcs{\hbox{$.\!\!^{\prime\prime}$}}
\def\farcm{\hbox{$.\!\!^{\prime}$}}

\def\nustar{{\sl NuSTAR}}
\def\chandra{{\sl Chandra}}
\def\cxo{{\sl CXO}}

\def\integral{{\sl INTEGRAL}}

\begin{document}

\title{A NuSTAR and Chandra Investigation of the Misaligned Outflow of PSR J1101--6101 and the Lighthouse Pulsar Wind Nebula}

\author{Noel Klingler}
\affil{Center for Space Sciences and Technology, University of Maryland, Baltimore County, Baltimore, MD, 21250, USA}
\affil{Astrophysics Science Division, NASA Goddard Space Flight Center, Greenbelt, MD, 20771, USA}
\affiliation{Center for Research and Exploration in Space Science and Technology, NASA Goddard Space Flight Center, Greenbelt, MD, 20771, USA}
\author{Jeremy Hare}
\affil{Astrophysics Science Division, NASA Goddard Space Flight Center, Greenbelt, MD, 20771, USA}
\affil{NASA Postdoctoral Program Fellow}
\author{Oleg Kargaltsev}
\affil{Department of Physics, The George Washington University, Washington, DC, 20052, USA}
\author{George G.\ Pavlov}
\affil{Department of Astronomy \& Astrophysics, The Pennsylvania State University, 525 Davey Laboratory, University Park, PA, 16802, USA}
\author{John Tomsick}
\affil{Space Sciences Laboratory, University of California, Berkeley, CA, 94720, USA}

\begin{abstract}
PSR J1101--6101 is an energetic young pulsar which powers the remarkable Lighthouse pulsar wind nebula (PWN).  
The pulsar belongs to the rare type of radio- and gamma-ray-quiet pulsars which are bright in hard X-rays.  
Moreover, the Lighthouse PWN is remarkable for its misaligned outflow (which gave rise to the PWN's nickname). 
Also known as ``pulsar filaments'', these collimated parsec-scale X-ray structures have been recently discovered in the vicinity of a handful of fast-moving pulsars, and appear unaffected by the ram pressure which confines pulsar tails.   
We report on \nustar\ observations of PSR J1101--6101 and its misaligned outflow -- the first observation of such a structure above $\sim$ 10 keV.  
We detect the outflow up to 25 keV, spatially resolve its spectral evolution with distance from the pulsar, find unambiguous evidence of spectral cooling with distance from the pulsar, and infer physical properties of the particles and magnetic field in the outflow.  
We also reanalzye archival \chandra\ data and discuss the outflow's small-scale structure. 
We detect pulsations from PSR J1101--6101 up to 20 keV, present the X-ray pulse profile, confirm its period derivative, and perform phase-resolved spectroscopy.
Lastly, we discuss the X-ray source 2CXO J110158.4--605649 = 2XMM J110158.5--605651 (a serendipitously observed blazar) and suggest it may be the X-ray counterpart to the GeV source 4FGL J1102.0--6054.
\end{abstract}


\section{INTRODUCTION}
Pulsars are one of nature's most powerful particle accelerators, capable of accelerating particles up to PeV energies.
As a pulsar rotates, it imparts its immense rotational-kinetic energy into a magnetized particle wind.
As the particles gyrate in the magnetic field, they emit synchrotron radiation (from radio to hard X-rays) and inverse-Compton radiation (in the MeV-GeV range) which can be seen as a pulsar wind nebula (PWN; see \citealt{Reynolds2017} for a recent review).

Pulsars receive birth kicks during their progenitor supernova explosions, and thus often travel with speeds on the order of hundreds of km s$^{-1}$ \citep{Verbunt2017}. 
If a pulsar moves through the interstellar medium (ISM) with supersonic speed, the ram pressure exerted by the ISM confines the pulsar wind into a parsec-long ``tail'' behind the moving pulsar \citep{Kargaltsev2017}.

A handful of fast-moving pulsars with observed tails are also accompanied by puzzling parsec-scale X-ray structures {\em strongly misaligned} with their pulsars' directions of motion (see, e.g.,  \citealt{vanEtten2008,Johnson2010,Pavan2014,Klingler2016a,Klingler2016b,Marelli2016,deVries2020,Klingler2020}). 
These so-called ``misaligned outflows'' (also referred to as ``pulsar filaments'') extend well beyond the extent of the bow shocks and appear to be unaffected by the ram pressure which confines the rest of the pulsar wind.
As these structures usually appear to be much brighter and longer on one side, it was proposed that they could be Doppler-boosted pulsar jets along the pulsar spin axis \citep{Pavan2014}, however, deep {\sl Chandra X-ray Observatory} (\cxo) images show that in at least three such supersonic PWNe (SPWNe), the jets remain confined to the bow shock interior and are unrelated to the misaligned outflows \citep{Klingler2016a,Klingler2016b}.

The parsec-scale sizes, lack of substantial bending, and hard spectra ($\Gamma=1.6-1.7$) exhibited by these outflows \citep{Kargaltsev2017} led to the suggestion that these structures could occur if high energy particles escaped into the ISM when the stand-off distance at the apex of the bow shock $r_s = (\dot{E}/4\pi c m_p n_{\rm ism} v_{\rm psr}^2)^{1/2}$, becomes smaller than the gyro-radius of high-energy electrons, $r_g=\gamma m_e c^2/eB_{\rm pwn}^{\rm apex}$ \citep{Bandiera2008}. 
In such a scenario, the escaped wind particles would follow the ambient ISM magnetic field lines (in terms of the bulk flow) and emit synchrotron radiation, therefore illuminating the ISM magnetic field structure.

The detection of a misaligned outflow produced by the transonic PSR J1809--1917 indicated that, in at least some SPWNe, alternative particle escape mechanisms can occur, as J1809's bow shock stand-off distance of 1 pc is incompatible with the compressed bow shock explanation \citep{Klingler2020}. 
Simulations by \citet{Barkov2019} and \citet{Olmi2019A,Olmi2019B} suggest that the particle escape can occur due to reconnection of the PWN and ISM magnetic fields in the bow shock (see also \citealt{Bykov2017}). 
In this scenario a magnetic bottle effect at the reconnection region can prevent lower-energy particles from escaping, which can explain why misaligned outflows are not seen in radio even when their accompanying PWNe are (in some cases).
Additionally, the opposing helicity of the magnetic fields in each PWN hemipshere causes magnetic reconnection to be favored in the hemisphere whose magnetic field direction opposes that of the ISM, which explains why these outflows are extremely asymmetrical, unlike typical pulsar jets (i.e., polar outflows).

Direct evidence for magnetic interaction is also seen in the Lighthouse PWN. 
High-resolution {\sl Chandra} images (see, e.g., Figure 15 of \citealt{Klingler2016a}) revealed ``magnetic draping'' \citep{Lyutikov2006,Dursi2008} of the ISM magnetic field (illuminated by the escaped pulsar wind) around the apex of the bow shock. 
Also, ATCA radio observations \citep{Pavan2014} showed that none (or very few) of the lower-energy (radio-emitting) pulsar wind particles present in the PWN tail escaped into the outflow, further suggesting either a magnetic bottle effect which screens out lower-energy particles and/or a reconnection region smaller than the gyro-radius of high-energy electrons.

It has also been proposed that misaligned pulsar outflows are the high-energy analogs of the nonthermal radio filaments seen toward the Galactic Center \citep{BarkovLyutikov2019}. 
However, observational evidence directly linking a radio filament and a pulsar misaligned outflow has yet to been seen. 
In contrast to the misaligned outflows associated with fast-moving pulsars, most of the Galactic Center filaments are only detected in radio and not in X-rays (only 4 of the 100+ radio filaments have X-ray counterparts; see, e.g., \citealt{YusefZadeh2021,Zhang2020}).

Only a handful of pulsar misaligned outflows are currently known.
Their broad-band spectral properties remain unknown as they are not seen in radio, and as no PWNe of this type have been observed above $\sim$10 keV. 
No substantial evidence of significant spectral cooling with distance from the pulsar has been found in a misaligned outflow so far. 
The degree of spectral cooling places constraints on particle energies, the particle propagation speed, and magnetic field strength inside the misaligned outflow.
This motivated us to obtain hard X-ray observations with \nustar\ of the brightest misaligned outflow, the Lighthouse PWN, powered by PSR J1101--6101, which we report in this paper.

The hard X-ray \integral\ source IGR J11014--6103 was first identified as a PWN by \citet{Pavan2011} based on the extended emission seen in X-rays with {\sl Swift}-XRT and {\sl XMM-Newton}.
\citet{Tomsick2012} proposed an association between this pulsar/PWN and the nearby supernova remnant (SNR) MSH 11--61A = G290.0--0.8 ($d=7\pm1$ kpc; \citealt{Reynoso2006}) based on the pulsar tail morphology and orientation (the tail points back toward the SNR), as resolved with {\sl Chandra} observations (Figure \ref{fig-lighthouse-PWN}, right panel).
This association and the estimated 10--30 kyr age of the SNR \citep{Garcia2012} imply a very high transverse velocity $v_\perp \sim800-2400$ km s$^{-1}$ (among the largest known pulsar velocities; assuming $d=7$ kpc). 
\citet{Halpern2014} detected pulsations with {\sl XMM-Newton}, revealing PSR J1101--6101 (henceforth J1101) with period $P=62.8$ ms, spin-down power $\dot{E}=1.6\times10^{36}$ erg s$^{-1}$, and characteristic age $\tau_c \equiv P/2\dot{P} = 116$ kyr. 
The pulsed fraction appears to increase with energy, reaching $\geq50$\% in 4--10 keV. 
Note that the association with the G290.0--0.8 SNR implies that the true age of the pulsar must be substantially younger than its characteristic age.

Deep \chandra\ observations (\citealt{Pavan2014,Pavan2016}; also see Figure \ref{fig-lighthouse-PWN}, right panel) revealed a hard pulsar spectrum ($\Gamma_{\rm psr}=1.08\pm0.08$, $F_{\rm 2-10\ keV}=6.2\times10^{-13}$ erg cm$^{-2}$ s$^{-1}$), a soft tail with $\Gamma_{\rm tail}=2.22\pm0.06$, $F_{\rm 2-10\ keV} = 6.1\times10^{-13}$ cgs), and a long ($>$5$'$) misaligned outflow ($\Gamma_{\rm mo}=1.7\pm0.1$, $F_{\rm 2-10\ keV}=6.5\times10^{-13}$ erg cm$^{-2}$ s$^{-1}$). 
The spectrum of the outflow appears to exhibit changes on scales $>$1$'$ (Figure 8 in \citealt{Pavan2016}).
The source is seen with \integral-ISGRI up to $60$ keV, but ISGRI cannot resolve the pulsar from the pulsar tail or misaligned outflow. 
J1101 is also one of only 7 pulsars with $\dot{E}>10^{36}$ erg s$^{-1}$ that are both radio and gamma-ray quiet (i.e., pulsations from these pulsars have only been detected in X-rays).

\begin{deluxetable}{lc}
\tablecolumns{9}
\tablecaption{Observed and Derived Pulsar Parameters \label{tbl-parameters}}
\tablewidth{0pt}
\tablehead{\colhead{Parameter} & \colhead{Value} }
\startdata
R.A. (J2000.0) & 11 01 44.96(9) \\
Decl. (J2000.0) & --61 01 39.6(7)  \\
Epoch of position (MJD) & 56,494  \\
Galactic longitude (deg) & 290.040  \\
Galactic latitude (deg) & --0.932  \\
Spin period, $P$ (ms) & 82.800077(6)  \\
Period derivative, $\dot{P}$ (10$^{-15}$) & $8.6(6)$ \\
Period epoch (MJD) & 56,494 \\
Distance, $d$ (kpc) & $7\pm1$  \\
Surface magnetic field, $B_s$ (10$^{11}$ G) & 7.4  \\
Spin-down power, $\dot{E}$ (10$^{36}$ erg s$^{-1}$) & 1.4  \\
Spin-down age, $\tau_{\rm sd} = P/(2\dot{P})$ (kyr) & 116 
\enddata
\tablenotetext{}{Parameters are from the ATNF Pulsar Catalog \citep{Manchester2005}.  The distance corresponds to that obtained from HI absorption measurements of SNR MSH 11--61A, by \citet{Reynoso2006}.}
\end{deluxetable}

\section{OBSERVATIONS \& DATA REDUCTION}

\subsection{NuSTAR}

\nustar\ \citep{2013ApJ...770..103H} observed J1101 on 2020 November 20 for 136 ks (ObsID 30601029002; PI: Klingler).
We reprocessed the data using {\tt HEASoft} v6.29c and \nustar\ CALDB v20211202 (see \citealt{Madsen2021} for details), which includes the latest calibration updates at the time of analysis/writing.
We ran the standard tool, {\tt nupipeline}, which applied all the latest calibrations and filtering, and barycenter-corrected the arrival times of photons originating from the pulsar's position, using \nustar\ clock correction file 20100101v128, which also corrects for \nustar's clock drift, providing a timing accuracy of $\sim$65 $\mu$s \citep{2021ApJ...908..184B}. 
We extracted spectra using {\tt nuproducts} (with option {\tt extended=yes} for analysis of the misaligned outflow).

\subsection{Chandra}
\chandra\ observed J1101 and its PWN for a total of 286 ks between Oct 2012 and Oct 2014 (PI: Pavan), with the ACIS-I detector (time resolution = 3.2 s).
We utilized the archival observations (see Table \ref{tbl-chandra}) in some of the spectral analyses of extended emission, described in subsequent sections.

\begin{deluxetable}{ccc}[h]
\tablecolumns{9}
\tablecaption{Archival \chandra\ Observations Used \label{tbl-chandra}}
\tablewidth{0pt}
\tablehead{\colhead{ObsID} & \colhead{Date} & \colhead{Exposure (ks)} }
\startdata
13787 & 2012 Oct 11 & 50 \\
16007 & 2014 Aug 28 & 116 \\
16517 & 2014 Sep 05 & 52 \\
17421 & 2014 Oct 02 & 20 \\
17422 & 2014 Oct 01 & 49
\enddata
\end{deluxetable}

We processed the data using the {\tt CIAO} software package v4.13 and CALDB v4.9.6.
We ran the standard tool {\tt chandra\_repro}, which applied all the latest calibrations and filtering, and extracted spectra using {\tt specextract} (with options {\tt weight=yes} and {\tt correctpsf=no} when analyzing regions of extended emission).
We restricted the \chandra\ data to the 0.5--8 keV energy range.

\begin{figure*}
\includegraphics[width=1.0\hsize,angle=0]{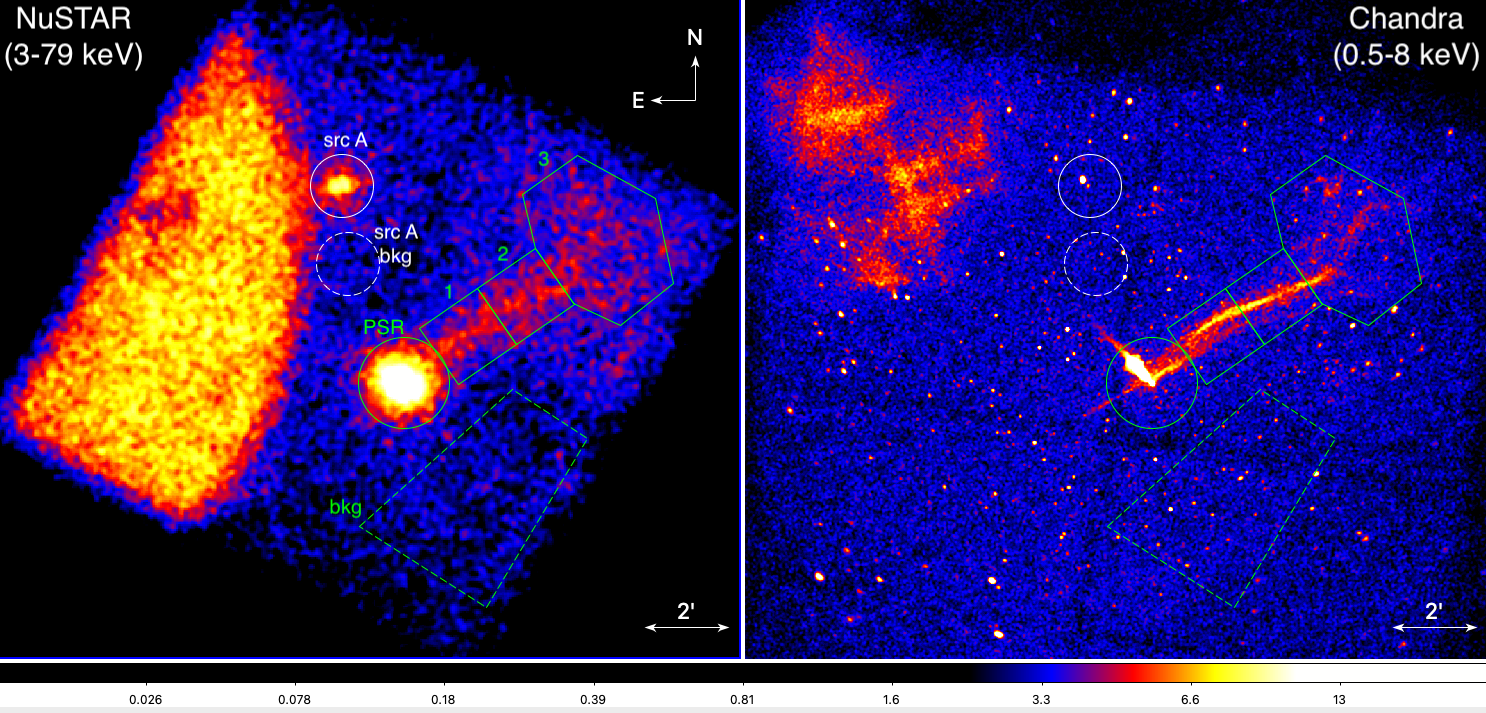}
\caption{Images of the J1101 field and the Lighthouse PWN.  {\sl Left:} \nustar\ (3--79 keV, FPMA+FPMB combined, 136 ks, smoothed with a 4-pixel Gaussian kernel).  {\sl Right}: {\sl Chandra} (ACIS-I, 0.5--8 keV, 286 ks, binned by a factor of 2 and smoothed with a 3-pixel ($r=3''$) Gaussian kernel).
The stray light seen in the Eastern part of the \nustar\ image is caused by nearby off-axis source (Cen X--3), and the diffuse emission seen in the Northeastern part of the {\sl Chandra} image is MSH 11--61A (PSR J1101's assumed progenitor SNR).
The color bar corresponds to the \nustar\ image and is in units of counts pixel$^{-1}$.
The following regions are shown:
PSR (green circle); outflow segments 1, 2, and 3 (green boxes/polygon); and source A (white circle).
The dashed polygonal region ``bkg'' was used for background subtraction for analysis of J1101 and the extended emission, and the dashed circle ``src A bkg'' was used only for analysis of source A (see \S3.3).}
\label{fig-lighthouse-PWN}
\end{figure*}

\subsection{Fermi-LAT}
No Fermi-LAT GeV source is reported at the position of PSR J1101 in the 4FGL--DR3 catalog, which was constructed using 12 years of Fermi-LAT data \citep{2022ApJS..260...53A}. 
We downloaded the Fermi-LAT data extending up to MJD 59771.6, which includes almost 14 years of observations, to search for a potential GeV counterpart to PSR J1101 or its PWN.  
The data spanned an energy range of 100 MeV to 300 GeV and included a 15$^{\circ}$ region of interest (ROI) centered on PSR J1101. The data were reduced using the {\tt Fermipy} software package \citep{2017ICRC...35..824W}, which uses the latest version (v2.2.0) of {\tt Fermitools}. 
We reduced the data following the standard procedures and used the {\tt P8R3\_SOURCE\_V3} version of the response function while selecting the ``Source'' event class (i.e., {\tt evclass=128}) and using both front and back converting events (i.e., {\tt evtype=3}). 
We then performed a binned analysis so that the energy dispersion correction could be used.

We first performed a search for a candidate counterpart to PSR J1101. 
To accomplish this, we used an initial model based on the source parameters from the 4FGL--DR3 catalog \citep{2022ApJS..260...53A}. 
The Galactic diffuse emission and isotropic emission were accounted for using the {\tt gll\_iem\_v07.fits} and {\tt iso\_P8R3\_SOURCE\_V3\_v1} models, respectively. 
We then freed the normalizations for all sources within 3 degrees of PSR J1101's position, and also allowed the normalization and photon index of the Galactic diffuse component to vary, as well as the normalization of the isotropic diffuse emission. 
Once the fit converged, we constructed a TS map of the region, assuming a power-law spectral model with a photon index of 2 for the test source, but no new source was found at the position of PSR J1101.
We then derived the 2$\sigma$ upper-limits on the Fermi-LAT flux of PSR J1101 from 0.3-3 and 3-30 GeV assuming a power-law spectral model with a photon index of 2: $6.6\times10^{-13}$ and $4.4\times10^{-13}$ erg cm$^{-2}$ s$^{-1}$.

\section{RESULTS}

\subsection{Misaligned Outflow}
\label{misaligned_out}
\begin{figure*}
\includegraphics[width=1.0\hsize,angle=0]{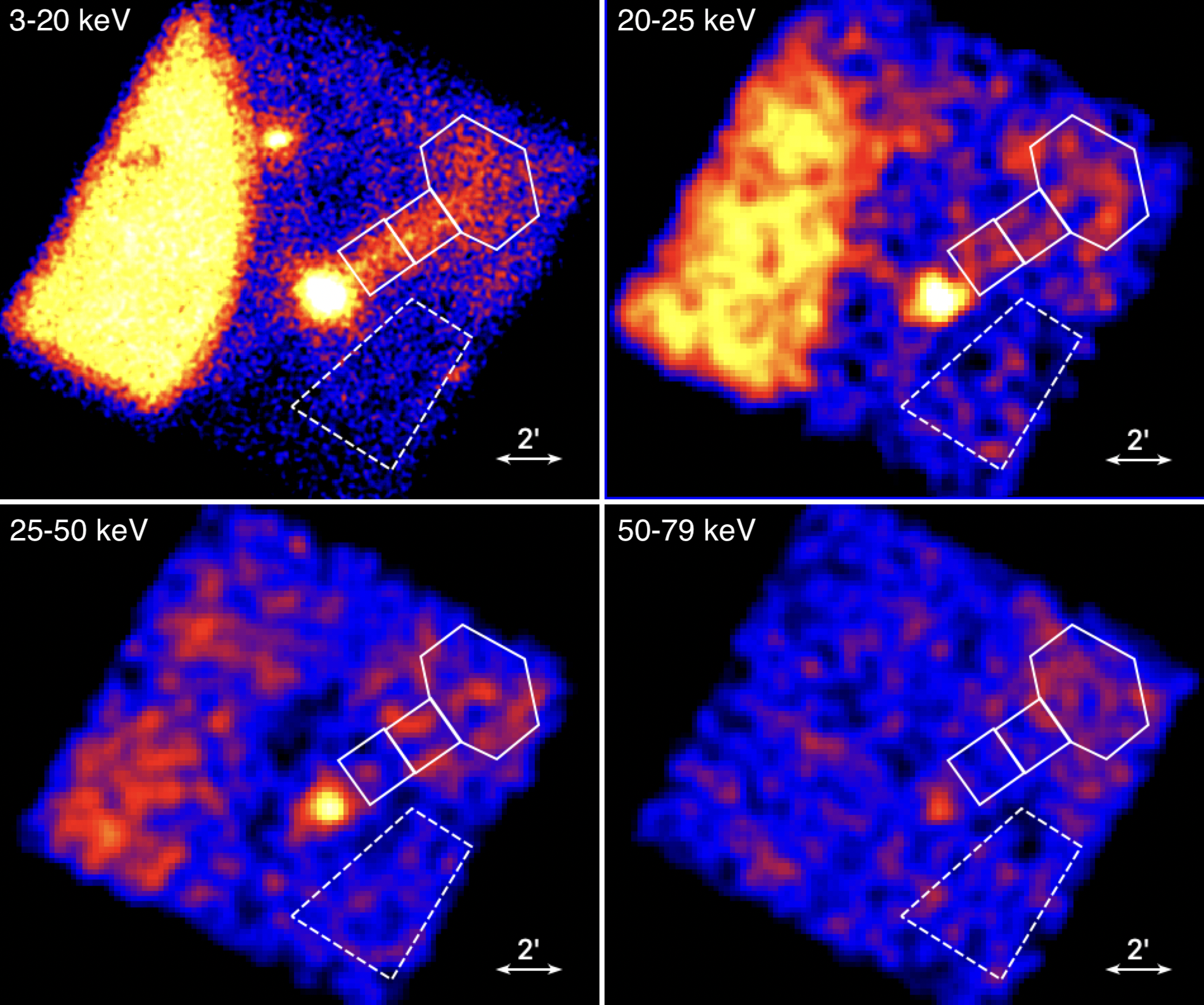}
\caption{\nustar\ images of the Lighthouse PWN in the 3--20 keV, 20--25 keV, 25--50 keV, and 50-79 keV bands (all images are smoothed with a 3-pixel Gaussian kernel).  The images of the latter 3 energy bands are binned by a factor of 4 for visual clarity.}
\label{fig-nustar-bands}
\end{figure*}

For the first time, we detected and resolved a pulsar misaligned outflow above $\sim$10 keV (i.e., above the \chandra\ and {\sl XMM} bands); we present the 3--79 keV image in Figure \ref{fig-lighthouse-PWN}. 
The outflow can be seen clearly and fully in the \nustar\ images up to energies $\sim$25 keV. 
Surprisingly, traces of segment 3 of the outflow can be faintly seen in the 50--79 keV band (Figure \ref{fig-nustar-bands}).
However this is likely due to elevated background emission in this vicinity (as is seen by the lower quality spectral fit to segment 3 compared to the two; see below).
The outflow is visible up to a distance of $\approx$ $7\farcm5$ from the pulsar; this is limited by the \nustar\ (and \chandra) FOV, so the outflow's actual extent may be longer.
The outflow's size and shape are comparable to those seen in the \chandra\ images, although the \nustar\ image is substantially blurred due to \nustar's broader PSF.
In the \nustar\ image we also see the outflow start to broaden and de-collimate about $5'$ away from the pulsar, as is also seen in the {\sl Chandra} image.

We extracted the outflow's spatially-resolved spectra using three different regions, shown by the green boxes and the polygon in Figure \ref{fig-lighthouse-PWN}.
We used a background region located ahead of the moving pulsar (i.e., southwest of it) to minimize contamination by  PWN emission.
We binned the spectra using {\tt ftgrouppha} and required a minimum S/N=6 for each bin. 
We fit the \nustar\ outflow spectra with an absorbed power-law (PL) model (XSPEC's {\tt tbabs}, which utilizes the
abundances of \citealt{Wilms2000}).
The absorbing Hydrogen column density was fixed to $N_{\rm H}=0.99\times10^{22}$ cm$^{-2}$, which is the best-fit value found by \citet{Pavan2016} while fitting both the pulsar tail and the outflow (independently) using \chandra\ data.
The model was multiplied by a constant which was set to 1.0 for the FPMA data and left as a free parameter for the FPMB data, in order to account for differing sensitivities between the two detectors.

\begin{figure}
\includegraphics[width=1.0\hsize,angle=0]{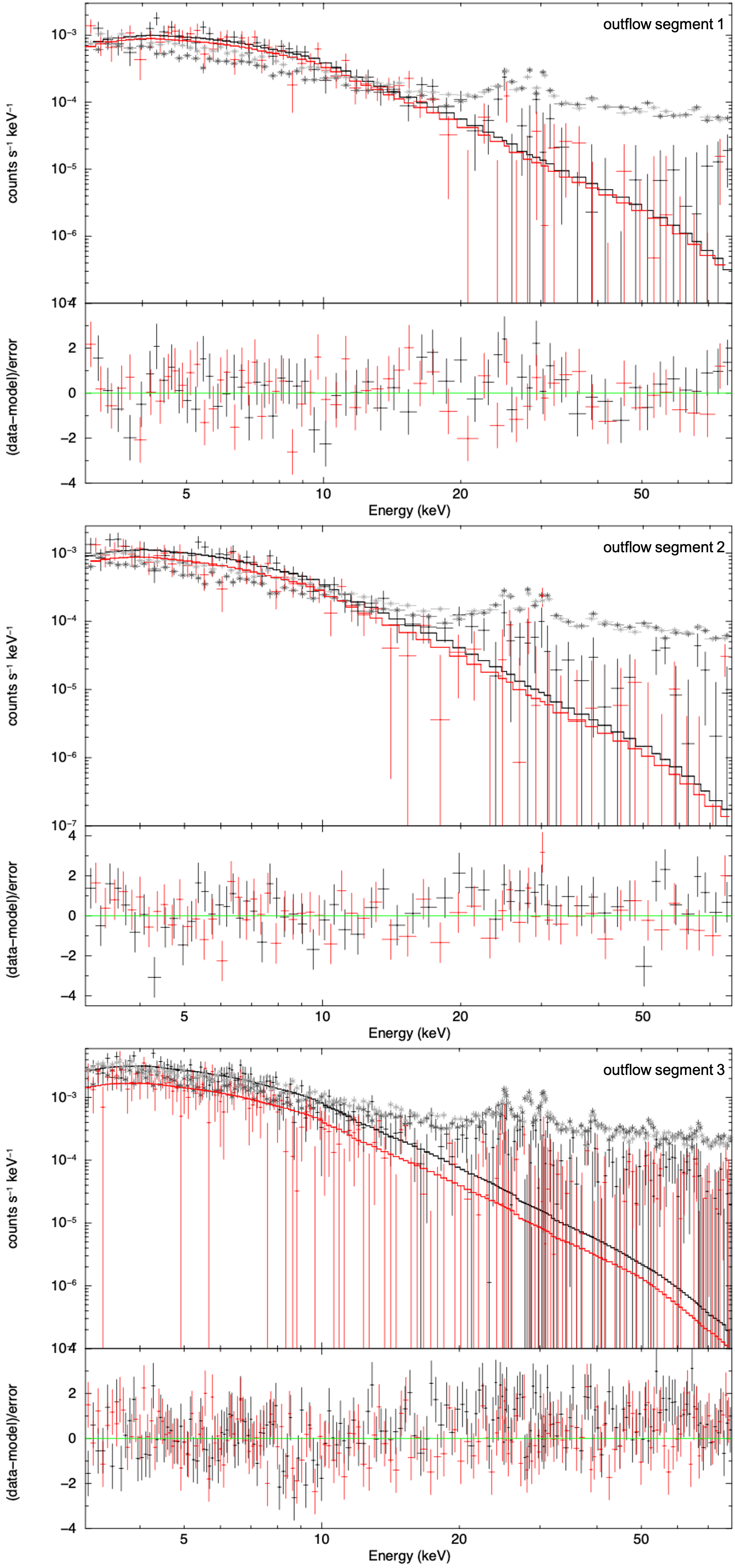}
\caption{Spectral fits of the three outflow segments to the \nustar\ data, binned by a minimum S/N=6.  The black and red data points show the FPMA and FPMB data, respectively, and the corresponding background is shown by the starred dark and light gray data points.  As mentioned in the text, restricting the fits to the 3--20 keV range does not noticeably change the best-fit spectral parameters.}
\label{fig-nustar-outflow-spectra}
\end{figure}

We show the fitted spectra in Figure \ref{fig-nustar-outflow-spectra} and list the fit parameters in Table \ref{table-outflow-spectra}.
Moving along the outflow (away from the pulsar), we found significant spectral evolution with distance from the pulsar, with the spectral slope changing from $\Gamma_1=1.79\pm0.08$ to $\Gamma_3=2.21\pm0.08$ (between segments 1 and 3, respectively; 3--79 keV).
To verify that our results were not skewed by the higher energy bins (in which the source count rates become progressively lower and in which the background becomes increasingly higher), we restricted the \nustar\ fits to the 3--20 keV range.
We obtained $\Gamma_1=1.81\pm0.08$, $\Gamma_2=2.08\pm0.09$, and $\Gamma_3=2.31\pm0.08$; thus, these results were consistent with the previous (3--79 keV) results. 
We note that the fit quality spectrum of segment 3 (reduced $\chi^2_{348}=1.21$) is slightly worse than that of segments 1 and 2 (reduced $\chi^2_{122}=1.04$ and $\chi^2_{122}=1.12$, respectively), which is likely due to the elevated background in the segment 3 vicinity (see the bottom right panel of Figure \ref{fig-nustar-bands}).

\begin{deluxetable*}{ccccccccc}
\tablecolumns{4}
\tablecaption{Misaligned Outflow \nustar\ Spectra}
\tablewidth{0pt}
\tablehead{\colhead{Segment} & \colhead{Area} & \colhead{Net Counts} & \colhead{$\Gamma$} & \colhead{$\mathcal{N}_{-5}$} & \colhead{Constant} & \colhead{$\chi_\nu^2$ (dof)} & \colhead{$F_{-13}$} & \colhead{$L_{33}$} }
\startdata
1 & 9,650 & 1,838$\pm74$ & $1.79\pm0.08$ & $7.6\pm1.2$ & $0.90\pm0.06$ & 1.04 (122) & $7.1\pm0.5$ & $4.2\pm0.3$ \\
2 & 9,660 & 1,863$\pm74$ & $2.03\pm0.09$ & $13.1\pm2.2$ & $0.81\pm0.06$ & 1.12 (122) & $6.3\pm0.6$ & $3.7\pm0.4$ \\
3 & 33,180 & 4,293$\pm147$ & $2.21\pm0.08$ & $64.3\pm9.5$ & $0.59\pm0.04$ & 1.21 (348) & $18.7\pm1.0$ & $11.3\pm0.6$ \\
\enddata
\tablenotetext{}{Spectral fit results for the misaligned outflow using \nustar\ data (3--79 keV).  Listed are the segment number, area (in units of arcsec$^2$), net counts, photon index $\Gamma$, PL normalization (in units of $10^{-5}$ photons s$^{-1}$ cm$^{-2}$ keV$^{-1}$ at 1 keV), the constant (fit to the FPMB data), reduced $\chi^2_{\nu}$ ($\nu$ degrees of freedom), observed (absorbed) 3-79 keV flux (in units of $10^{-13}$ erg cm$^{-2}$ s$^{-1}$), and luminosity (in units of $10^{33}$ erg s$^{-1}$, assuming $d=7$ kpc).  In all fits we set $N_{\rm H}=0.99\times10^{22}$ cm$^{-2}$.
}
\label{table-outflow-spectra}
\end{deluxetable*}

For a comparison, we also extracted and fit the spectra of the three outflow segments from the \chandra\ data (0.5--8 keV).
We excluded all point sources present in the source and background regions, and binned the spectra by requiring S/N=4 for each bin; the results are listed in Table \ref{table-outflow-chandra-spectra}.
For segment 1, the spectra are compatible: $\Gamma_{\rm 1,CXO}=1.71\pm0.04$ and $\Gamma_{\rm 1,Nu}=1.79\pm0.08$, although in the subsequent segments the differences become prominent: $\Gamma_{\rm 2,CXO}=1.86\pm0.05$ versus $\Gamma_{\rm 2,Nu}=2.03\pm0.09$, and $\Gamma_{\rm 3,CXO}=1.74\pm0.05$ versus $\Gamma_{\rm 3,Nu}=2.21\pm0.08$.
Thus, the \chandra\ spectra do not show consistent spectral cooling with distance contrary to the \nustar\ data.
This might suggest that the effects of spectral cooling in the outflow are only noticeably seen at energies above 8 keV.

As similar differences were also seen in the comparison of \chandra\ and \nustar\ data in studies of the high-mass X-ray/gamma-ray binary LS 5039 \citep{Volkov2021} and PSR J1617--5055 \citep{Hare2021}, they are likely due to calibration uncertainties (see also \citealt{Madsen2017}).
Therefore, joint \chandra+\nustar\ fits of the Lighthouse data are not warranted here.

\begin{deluxetable*}{cccccccc}
\tablecolumns{4}
\tablecaption{Misaligned Outflow \chandra\ Spectra}
\tablewidth{0pt}
\tablehead{\colhead{Segment} & \colhead{Area} & \colhead{Net Counts} & \colhead{$\Gamma$} & \colhead{$\mathcal{N}_{-5}$} & \colhead{$\chi_\nu^2$ (dof)} & \colhead{$F_{-13}$} & \colhead{$L_{33}$} 
}
\startdata
1 & 9,650 & 4,629$\pm92$ & $1.71\pm0.04$ & $7.4\pm0.4$ &  1.12 (185) & $2.6\pm0.1$ & $2.4\pm0.1$ \\
2 & 9,660 & 4,146$\pm90$ & $1.86\pm0.05$ & $8.5\pm0.4$ & 0.99 (174) & $2.4\pm0.1$ & $2.5\pm0.1$ \\
3 & 33,180 & 5,803$\pm157$ & $1.74\pm0.05$ & $12.7\pm0.6$ & 1.02 (389) & $4.2\pm0.1$ & $4.0\pm0.1$ \\
\enddata
\tablenotetext{}{Spectral fit results for the misaligned outflow using \chandra\ data (0.5--8 keV).  Listed are the segment number, area (in arcsec$^{2}$), net counts, photon index $\Gamma$, PL normalization (in units of $10^{-5}$ photons s$^{-1}$ cm$^{-2}$ keV$^{-1}$ at 1 keV), reduced $\chi^2_{\nu}$ ($\nu$ degrees of freedom), observed (absorbed) 0.5--8 keV flux (in units of $10^{-13}$ erg cm$^{-2}$ s$^{-1}$), and luminosity (in units of $10^{33}$ erg s$^{-1}$, assuming $d=7$ kpc).  In all fits we set $N_{\rm H}=0.99\times10^{22}$ cm$^{-2}$.}
\label{table-outflow-chandra-spectra}
\end{deluxetable*}

\subsection{Pulsar}

\subsubsection{Timing}
\label{timing}

Up until recently, J1101 had no long-term timing solution as it is neither a radio nor a gamma-ray pulsar. 
However, \citet{Ho2022} used {\sl NICER} monitoring to obtain an updated timing solution for J1101, which overlaps with the epoch of the \nustar\ observation. 
This timing solution was published shortly before the conclusion of this work, so we used the pulse period that we found in the \nustar\ data to create phase folded light curves and phase-resolved spectra. 
To investigate the dependence of the pulse profile on energy, we used only the \nustar\ data.
To find the spin period and its uncertainty, we used the $Z_1^2$ test \citep{1983A&A...128..245B}. 
Given that \nustar\ cannot resolve the extended PWN from the point-like pulsar, we searched over different source extraction aperture radii and photon energies to maximize the signal. 
We found that an extraction radius of 60$''$ ($\approx$75\% PSF) and photons in the 3--40 keV energy range give the maximum $Z^2_1$.

After applying the optimal energy and source extraction radius cuts, we found a $Z^2_1=208.6$ peak at a frequency of $f=15.92303947(15)$ Hz. 
To estimate the 1$\sigma$ uncertainty on the frequency, we use $\sigma_f=(\sqrt{3}/\pi)T^{-1}_{\rm span}(Z^2_{\rm 1,max})^{-1/2}$, which is valid when the pulsations are purely sinusoidal (i.e., $m=1$) and there are no gaps in the time series (see, e.g., \citealt{Hare2021}). 
Of course, the J1101 time series has gaps due to the source being occulted by the Earth throughout \nustar's orbit, so this formula only provides an estimate on the uncertainty. We find that this timing solution at the epoch $t_0=59175.34321456$ MJD is consistent within uncertainties with the timing solution recently reported by  \cite{Ho2022}.

We present the 3--40 keV pulse profile for J1101 in Figure \ref{pulse_profile}. 
The pulse profile shows a single wide pulse with a relatively broad flat top without a sharp peak. 
As previously mentioned, \nustar\ cannot resolve the pulsar from the PWN and the part of the outflow in the pulsar's vicinity, which makes it difficult to accurately estimate the pulsed fraction (or explore its dependence on energy) as we do not reliably know the contribution from those structures. 
Modeling the contribution from the tail and misaligned outflow using spectra extracted from the  \chandra\ data would not yield reliable results due to the fact that the same regions exhibit different spectra in the \chandra\ and \nustar\ data due to the calibration uncertainties, as demonstrated above in \S3.1.

\subsubsection{Phase-Integrated Spectroscopy}
\label{phase_int_spec}
We first performed fits to the phase integrated pulsar spectra. To do this, we extract the spectra from a $r=$30$''$ circular region centered on the pulsar. 
\nustar\ is unable to resolve the pulsar from the PWN, which has an extent of $\sim30''$ for the brightest part (see  Figure 2 in \citealt{Pavan2016}), but the PWN can be roughly modeled when modeling the pulsar's spectrum. 
Modeling the spectrum of the misaligned outflow is much more difficult as it is unclear what fraction of the outflow is contained in a given region. 
Therefore, this extraction region was chosen to minimize the impact of the contamination of the emission from the base of the misaligned outflow on the pulsar's spectrum. 
The background was extracted from a source-free region offset from the pulsar, PWN, and misaligned outflow. Prior to fitting, the spectra were binned to have a signal-to-noise ratio of at least 5 per bin. 
The spectra from the two focal plane modules (FPMA and FPMB) were simultaneously fit, and we use a {\tt const} parameter to account for calibration uncertainties between the two detectors. We find that the difference between normalizations remains $<5\%$ throughout all fits of the pulsar$+$PWN spectra.

We fit the spectra with an absorbed power-law model. 
The absorbing column density was frozen to the same value used to fit the misaligned outflow (i.e., $N_{\rm H}=0.99\times10^{22}$ cm$^{-2}$; see Section \ref{misaligned_out}). 
We find a photon index $\Gamma=1.48\pm0.03$, an unabsorbed 3--79 keV flux of (5.2$\pm0.2)\times10^{-12}$ erg cm$^{-2}$ s$^{-1}$, and $\chi^2=148$ for 149 degrees of freedom (dof). 
This photon index is larger than previously found by \citet{Pavan2016} for the pulsar using \chandra\ data ($\Gamma_{\rm CXO}=1.08\pm0.08$), most likely due to the PWN which was not accounted for in our fits. 
To account for this PWN emission we use the results of the spatially resolved spectral fits performed by \citet{Pavan2016}. 
To accomplish this, we included the anticipated contribution of the PWN as an additional power-law component in the model, freezing the flux and photon index to the values of $\Gamma=2.22\pm0.06$ and $F_{\rm 2-10  \ keV}=6.1\times10^{-13}$ erg cm$^{-2}$ s$^{-1}$ found by \citet{Pavan2016}.  
After accounting for the PWN emission we find that the pulsar's fitted photon index decreases to $\Gamma=1.14\pm0.04$ and the unabsorbed 3--79 keV flux remains virtually the same: (4.9$\pm0.3)\times10^{-12}$ erg cm$^{-2}$ s$^{-1}$, with $\chi^2=146$ for 149 d.o.f. 
The pulsar's flux in the 2--10 keV band, $F_{\rm 2-10 \  keV}=(6.6\pm0.3)\times10^{-13}$ erg cm$^{-2}$ s$^{-1}$ is comparable to that found by \citet{Pavan2016}.

We also tried a larger region (i.e., $r=50''$) to increase the statistics and better constrain the photon index.  However, we found that, after accounting for the PWN emission, the photon index becomes much larger ($\Gamma=1.31\pm0.03$), and the flux, $F_{\rm 2-10 \  keV}=(9.2\pm0.3)\times10^{-13}$ erg cm$^{-2}$ s$^{-1}$, increases, becoming incompatible with the pulsar flux reported by \citet{Pavan2016}. 
This increase in flux and the larger photon index are likely due to the additional contribution from the base of the misaligned outflow contaminating the pulsar$+$PWN spectrum. 
This further supports the choice of a smaller spectral extraction region.

\subsubsection{Phase-Resolved Spectroscopy}

\begin{figure}
\includegraphics[width=1.0\hsize,angle=0]{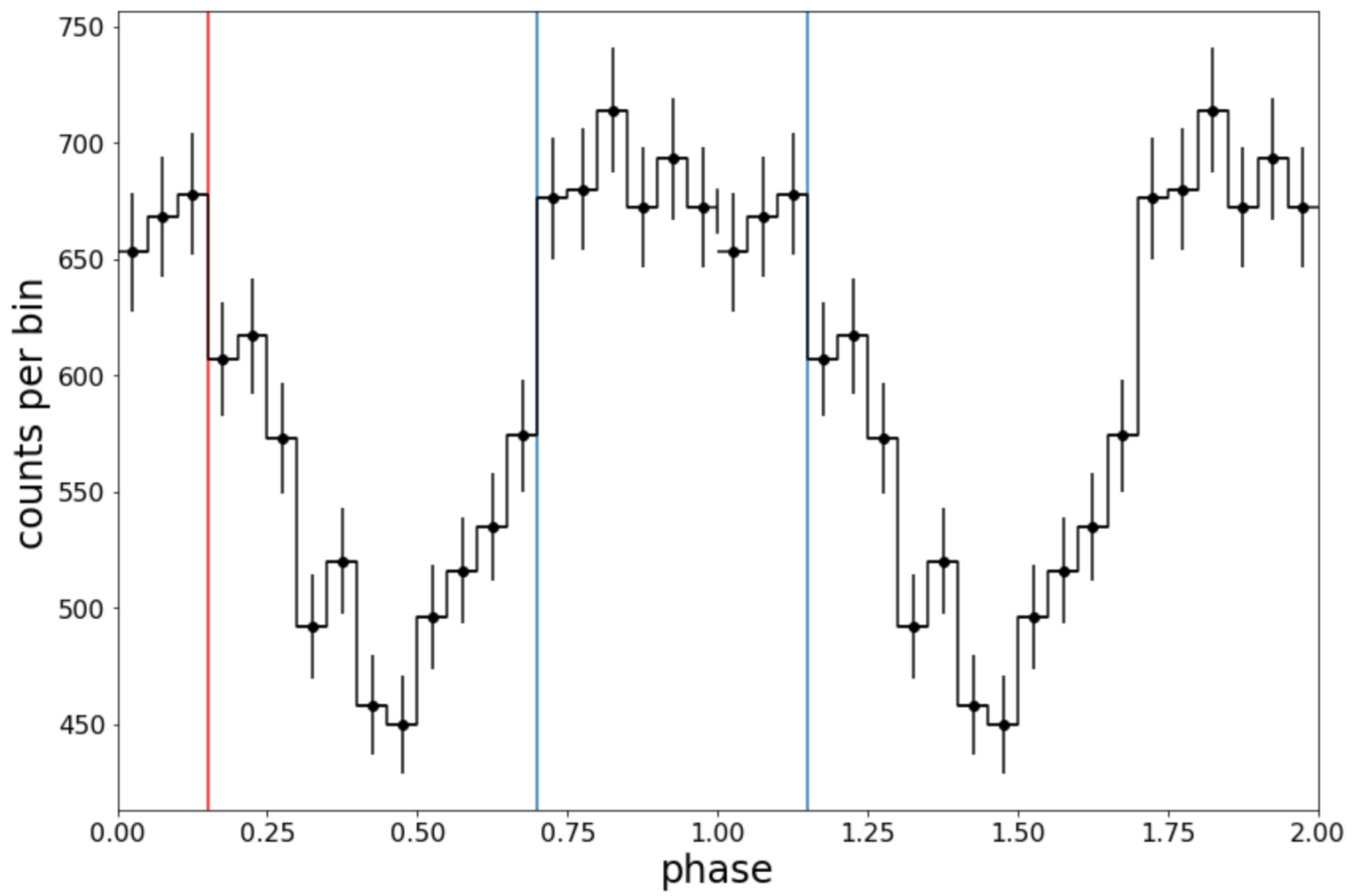}
\caption{J1101 pulse profile in the 3-40 keV energy range. The red line shows the start of the phase range used for pulse minimum, which extends to the first blue line. Beyond the first blue line (at $\phi=0.70$) up until the second blue line corresponds to the phase range used for pulse maximum.}
\label{pulse_profile}
\end{figure}

In addition to the phase-integrated spectral fits we have also attempted phase-resolved spectroscopy. 
The spectra were extracted from the same regions discussed in Section \ref{phase_int_spec}, and the absorbing column density was frozen to the same value. We chose two phase bins corresponding to the pulse maximum (from $0<\phi_{\rm max}<0.15$ and $0.7<\phi_{\rm max}<1.0$) and minimum ($0.15<\phi_{\rm min}<0.70$; see Figure \ref{pulse_profile}).
Fitting the spectra from both phase ranges with an absorbed PL model led to differing photon indices of 
$\Gamma_{\rm max}=1.41\pm0.04$  and $\Gamma_{\rm min}=1.58\pm0.05$ for the pulse maximum and minimum, respectively. 
However, in these fits we have again neglected the contribution from the PWN, which is not resolved by \nustar\ and which has a similar 2--10 keV flux to the pulsar (see Section \ref{phase_int_spec}). 

We account for the PWN emission for both phase ranges in the same way as discussed in Section \ref{phase_int_spec}. 
The resulting fitted values for the photon indices were $\Gamma_{\rm max}=1.11\pm0.05$ ($\chi^2_{85} = 0.91$) and $\Gamma_{\rm min}=1.18\pm0.07$ $\chi^2_{70}=0.84$).
These results suggest that we cannot measure the difference between pulse maximum and pulse minimum when accounting for the PWN emission. 
The 3--79 keV unabsorbed fluxes for the pulse maximum and minimum (after accounting for the PWN emission) are $F_{\rm max}=(6.8\pm0.5)\times10^{-12}$ erg cm$^{-2}$ s$^{-1}$ and $F_{\rm min}=(3.8\pm0.4)\times10^{-12}$ erg cm$^{-2}$ s$^{-1}$, respectively.

\subsection{Source A = 2CXO J110158.4--605649: A Possible Counterpart to 4FGL J1102.0--6054}
A serendipitously-detected field point source, ``Source A'', is also visible in the \nustar\ image at approximately R.A., decl.\ = 11:01:57.5, --60:56:58.3 (see Figure \ref{fig-lighthouse-PWN}).
SIMBAD lists two entries within 15$''$ of Source A's position: the X-ray source 2XMM J110158.5--605651 (which is the bright source seen at Source A's position in the {\sl CXO} image; Figure \ref{fig-lighthouse-PWN} right panel), and the radio source RJG2006 F \citep{Reynoso2006}.
These two sources are positionally consistent with each other (within 5$''$) and with 2CXO J110158.4--605649 (R.A., decl\ = 11:01:58.46, --60:56:49.91; $\pm0\farcs74$, 95\% CL) and are thus likely counterparts to the same object as well as  \nustar\ Source A.
For RJG2006 F, \citet{Reynoso2006} list a radio spectral index $\alpha = 0.5\pm0.1$, a 20-cm spectral flux density $S=57$ mJy (where $S\propto \nu^\alpha$), a systemic velocity $>80$ km s$^{-1}$ (measured from its HI emission), and a kinematic distance $>14$ kpc.
There is a Gaia source (source ID 5337957089788123136; parallax distance $d=4.34$ kpc) located 1\farcs95 from the \chandra\ position, though given its positional offset and its Galactic nature (in contrast to the likely extragalactic nature of Source A; see below), it is likely unrelated.

To find the $N_{\rm H}$ of Source A (=2CXO J110158.4--605649), we first fitted the \chandra\ data and found that its spectrum can be described by an absorbed PL model with $N_{\rm H}=(4.46\pm0.67)\times10^{22}$ cm$^{-2}$, $\Gamma=1.43\pm0.20$,  $\mathcal{N}=(5.1\pm1.7)\times10^{-5}$ photon s$^{-1}$ cm$^{-2}$ keV$^{-1}$ (at 1 keV), and $\chi_{119}^2=0.92$.
The best-fit $N_{\rm H}$ value is well in excess of the maximum Galactic $N_{\rm H}$: HEASARC's $N_H$ tool\footnote{https://heasarc.gsfc.nasa.gov/cgi-bin/Tools/w3nh/w3nh.pl} lists the maximum Galactic $N_{\rm H}$ in the vicinity ($r<5'$) of Source A as being in the range of $N_{\rm H} = (1.17 - 1.26)\times 10^{22}$ cm$^{-2}$.

Next we fitted the \nustar\ data (independently from the \chandra\ data) with both models, and set $N_{\rm H}$ to the best-fit values (as found by \chandra) for each model.
Both models provided satisfactory fits, with best-fit parameters consistent with those found by \chandra.
For the absorbed PL model, the \nustar\ data yielded $\Gamma=1.70\pm0.10$, $\mathcal{N}=(10.5\pm2.3)\times10^{-5}$ photons s$^{-1}$ cm$^{-2}$ keV$^{-1}$ at 1 keV, and
$\chi^2_{209}=0.98$.
Considering Source A = 2CXO J110158.4--605649's extragalactic nature implied by its $N_{\rm H}$ and HI emission, we propose that it is an AGN.

\begin{figure*}
\includegraphics[width=1.0\hsize,angle=0]{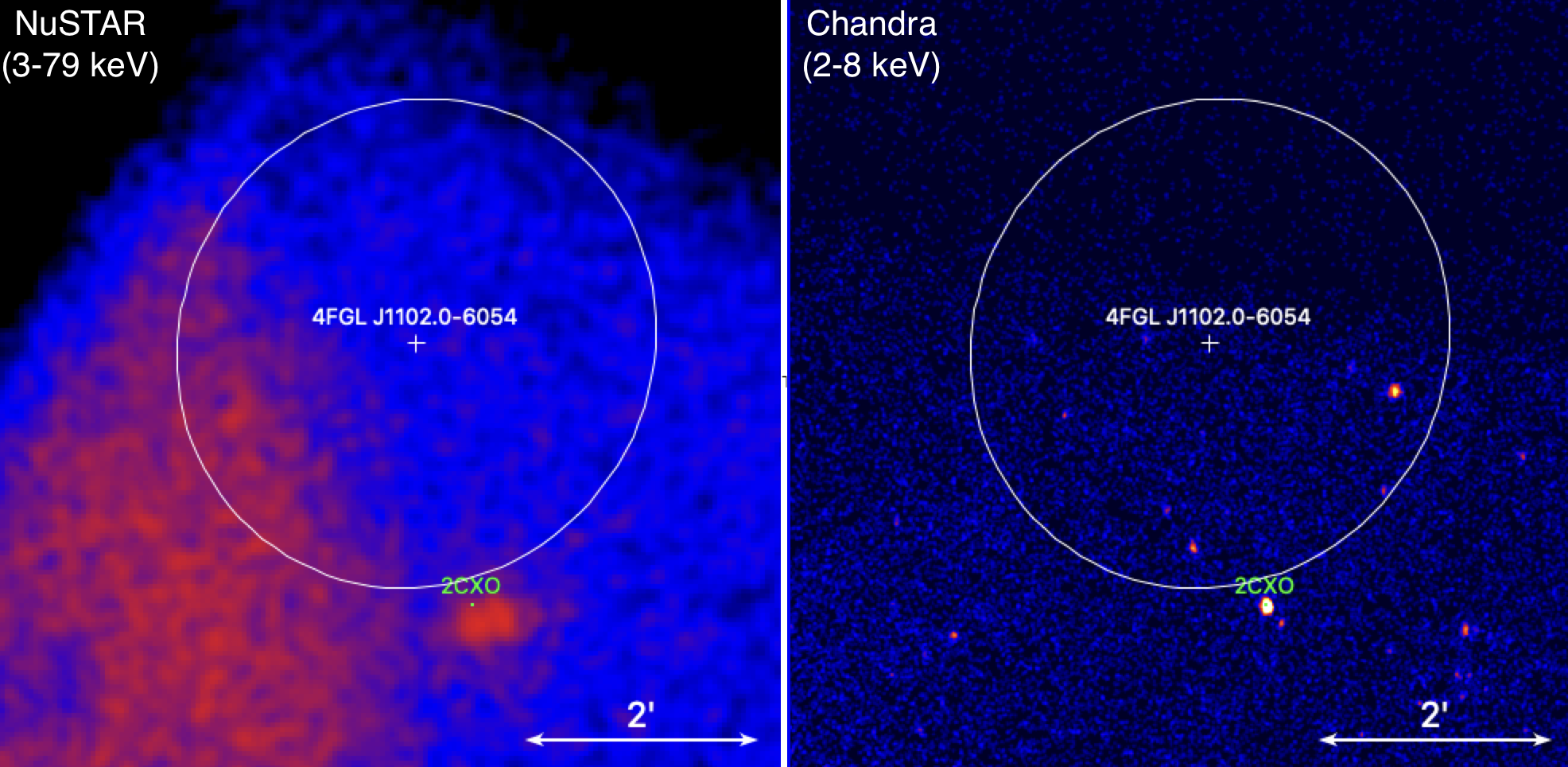}
\caption{\nustar\ and \chandra\ images of the vicinity of 4FGL J1102.0-6054 (95\% error ellipse is shown).  Also shown is Source A = 2CXO J110158.4--605649 (abbreviated as ``2CXO'').  The \chandra\ image was restricted to the 2--8 keV range in attempt to filter out possible thermal emission from nearby stars (i.e., sources unrelated to 4FGL J1102).}
\label{fig-src_A}
\end{figure*}

\begin{figure}
\includegraphics[width=1.0\hsize,angle=0]{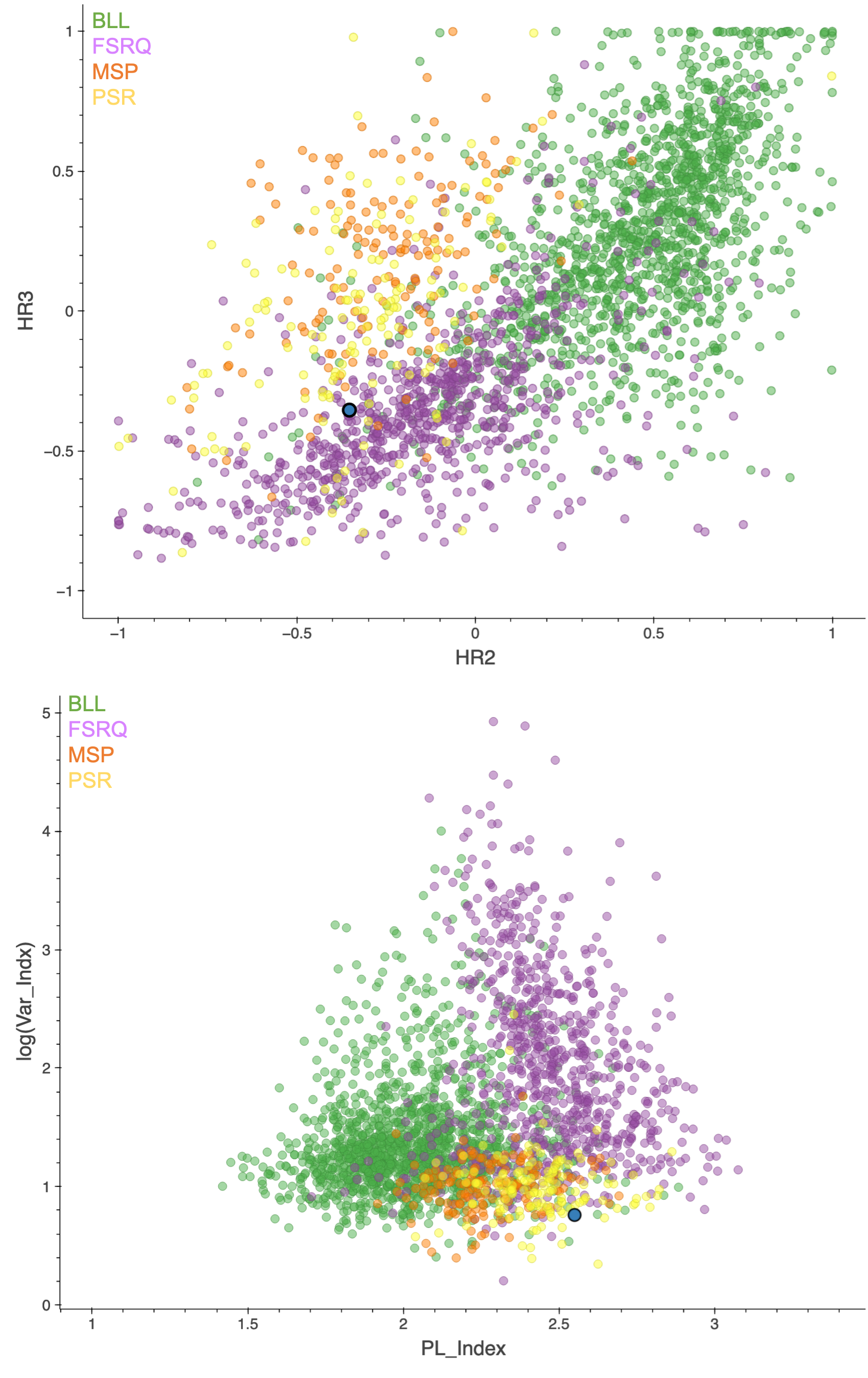}
\caption{Plots of 4FGL--DR3 sources in different phase spaces.  Shown are the most populous source classes (besides unidentified sources): BL Lac objects (BLL; green), flat spectrum radio quasars (FSRQs; magenta), millisecond pulsars (MSPs; orange), and pulsars (PSRs; yellow); the blue bolded point marks 4FGL J1102.0--6054. 
{\sl Top:} Hardness ratios HR2 vs HR3 (where HR2 = (F7+F6+F5--F4)/(F7+F6+F5+F4),  HR3 = (F6+F5+F4--F2--F3)/(F6+F5+F4+F2+F3), and $F\#$ are the fluxes in different {\sl Fermi}-LAT energy bands as defined in 4FGL--DR3. 
{\sl Bottom:} Log of the variability index vs the PL index.
Plots were produced by GCLASS:  \url{https://home.gwu.edu/~kargaltsev/GCLASS}
}
\label{fig_gclass}
\end{figure}

Our \nustar\ pointing also serendipitously covered the 95\% error ellipse of 4FGL J1102.0--6054 (as well as the archival \chandra\ observations; see Figure \ref{fig-src_A}).
4FGL J1102.0--6054 is classified (based on its spectral properties) as a blazar candidate of uncertain type (BCU) in the 2nd Data Release (DR2) of the 4th {\sl Fermi}-LAT Catalog (\citealt{Ballet2020}; the source was not listed as having its classification updated in DR3; \citealt{Fermi2022}). 
In Figure \ref{fig_gclass} we plot the spectral characteristics of the four most populous 4FGL classes of identified 4FGL sources (which account for 89.2\% of identified sources), and 4FGL J1102.
4FGL's hardness ratios, variability index, and PL index appear compatible with the range of those exhibited by flat spectrum radio quasars (FSRQs).
Thus, with Source A's extragalactic nature (implied by its radio properties and X-ray spectrum), with its PL spectrum being compatible with AGN emission, and with the absence of any other bright hard X-ray sources in the vicinity, we propose an association with 4FGL J1102 and a support its classification as a blazar (FSRQ).

\section{Discussion}

\subsection{Misaligned Outflow}

\subsubsection{Small-Scale Structure}

\begin{figure*}
\includegraphics[width=1.0\hsize,angle=0]{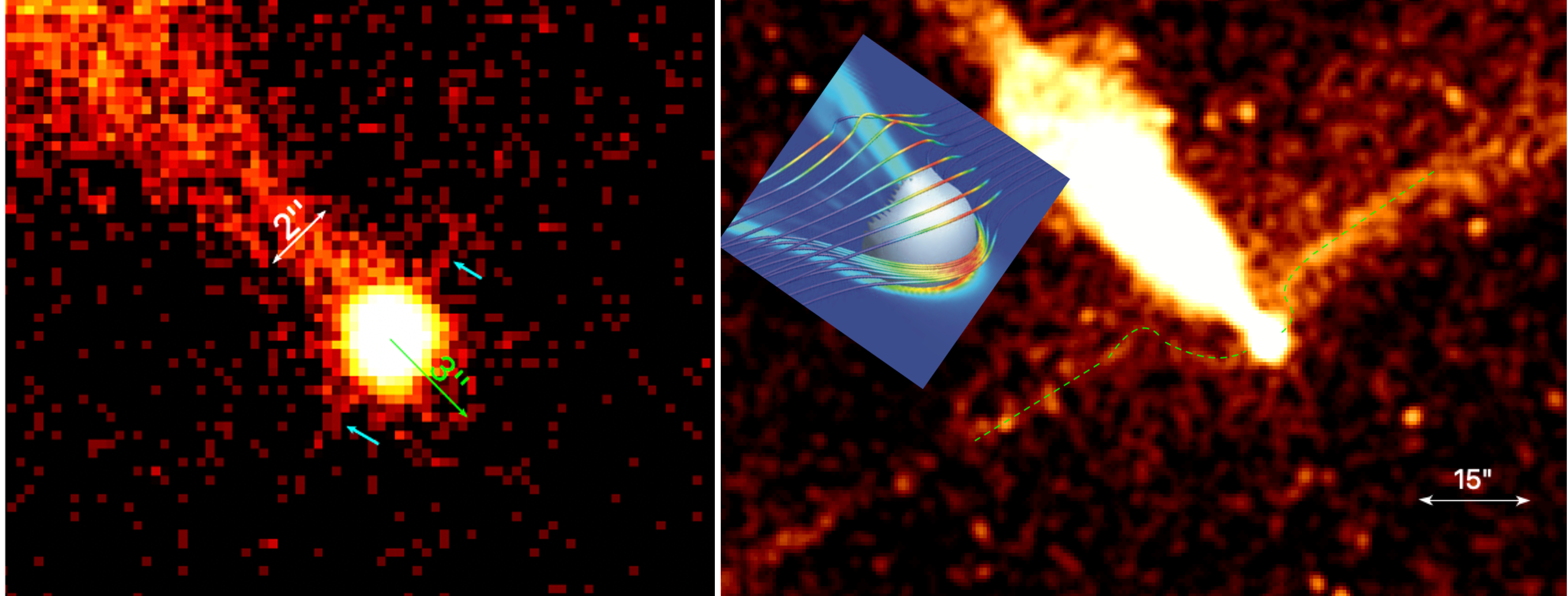}
\caption{\chandra\ images showing the fine structure in the vicinity of the Lighthouse PWN bow shock.  {\sl Left}: Zoomed-in image of the pulsar (binned by a factor of 0.5 to show sub-arcsecond features) showing hints of narrow streams originating from the pulsar (marked by cyan arrows), and enhanced emission seen up to 3$''$ ahead of the pulsar (shown by the green arrow).  {\sl Right:} Zoomed-out image showing evidence of ``magnetic draping'' of the ISM magnetic field lines around the PWN bow shock (marked by the green dashed curves).
The inset image is an illustration of magnetic draping from \citet{Dursi2008}.}
\label{fig-fine-structure}
\end{figure*}

To investigate the misaligned outflow's small-scale morphological features, we reanalyzed the high-resolution \chandra\ data.
In Figure \ref{fig-fine-structure} we present the merged \chandra\ image binned by a factor of 0.5 to show sub-arcsecond features. 
There appears to be a region of faint emission which extends for $\approx 2''$--3$''$ ahead of the pulsar (shown by the green arrow in the left panel of Figure \ref{fig-fine-structure}). 
This distance significantly exceeds a plausible projected bow shock standoff distance, $\theta_s\simeq0\farcs34\,(\mu/15\, {\rm mas\ yr^{-1}})n_{\rm H}^{-1/2} (d/7~{\rm kpc})^{-2} \sin i$, where $\mu$ is the pulsar's proper motion, $n_{\rm H}$ is the number density of the ambient medium (in units of cm$^{-3}$), and $i$ is the inclination angle of the pulsar's velocity with respect to the line of sight (see, e.g., \citealt{Brownsberger2014}). 
The enhancement of brightness slightly ahead of the pulsar can be explained by the escape of particles (likely facilitated by magnetic reconnection; further explained below) into the region of the draped and compressed magnetic field, illustrated by the inset in Figure \ref{fig-fine-structure}.

Figure \ref{fig-fine-structure} (left panel) also shows two narrow streams originating from the pulsar region (marked by the cyan arrows).
These streams may originate in opposite halves of the bow shock (see \citealt{Barkov2019,Olmi2019A} for a discussion). 
These streams, clearly visible in the 0.5--5 keV range, are composed of particles that escape the bow shock apex region (which is unresolved from the pulsar in the X-ray images). 
We note that the pulsar wind particle escape requirement of having the gyroradius exceeding the bow shock stand-off distance, $r_g\gtrsim r_s= \theta_s d/\sin i$ (originally proposed by \citealt{Bandiera2008} for the Guitar PWN), translates into a rather stringent upper limit on the post-shock magnetic field $B\lesssim 5 (\mu/15~{\rm mas~yr}^{-1})^{-2/3}(d/7~{\rm kpc})^{2/3}(E/1~{\rm keV})^{1/3}n_H^{1/3}$~$\mu$G, where $E$ is the synchrotron photon energy.
Since this field cannot be lower than the ISM field, $B \gtrsim B_{\rm ISM} \sim 5$ ${\rm \mu G}$ \citep{Ferriere2015}, the synchrotron emission can only be expected above $E_c\sim 1 (B_{\rm ISM}/5~\mu{\rm G})^3 (\mu/15~{\rm mas~yr}^{-1})^{2}(d/7~{\rm kpc})^{-2}n_H^{-1}$ keV because particles emitting synchrotron radiation at lower energies should not be escaping as per the above requirement. 
Sensitive observations at IR frequencies can therefore test the \cite{Bandiera2008} escape condition and place constraints on $B_{\rm ISM}$, $n_H$, and the distance to the pulsar. 
If the outflow is detected at frequencies far below those of X-rays (e.g., IR), it may imply that the particles in the outflow are ISM particles accelerated in the forward shock region \citep{Bykov2017}.

Particle escape can also be facilitated by magnetic reconnection of the external (ISM) magnetic field with the PWN's magnetic field \citep{Barkov2019}.
This is supported by the asymmetry of the outflow (i.e., the outflow not having a comparable counterpart on the opposite side of the PWN) because in the magnetic reconnection scenario, the reconnection leads to particle acceleration and escape on the side of the PWN where the PWN magnetic field is directed opposite the ISM magnetic field (see Figure 5 of \citealt{deVries2022} for an illustrative diagram). 
In the context of this scenario, we note that it is puzzling why the two acrsecond-scale mini-jets (shown in the left panel of Figure \ref{fig-fine-structure}) appear to be similar in brightness and size while the western outflow becomes far more prominent than the eastern one on larger scales and at greater distances from the pulsar (see Figure \ref{fig-fine-structure}, right panel).

The deep \chandra\ images also show that (on arcsecond scales) the brightest part of the outflow (i.e., near the pulsar) does not just stream out along a straight line from the pulsar. 
In the vicinity of the pulsar the outflow initially bends back in the direction opposite of the pulsar's motion (toward the northeast) but then sharply turns toward the northwest (see the dashed green curve in the right panel of Figure \ref{fig-fine-structure}).
This indicates the presence of magnetic ``draping'' \citep{Lyutikov2006,Dursi2008} of the ISM magnetic field lines around the PWN bow shock.

\begin{figure*}
\includegraphics[width=1.0\hsize,angle=0]{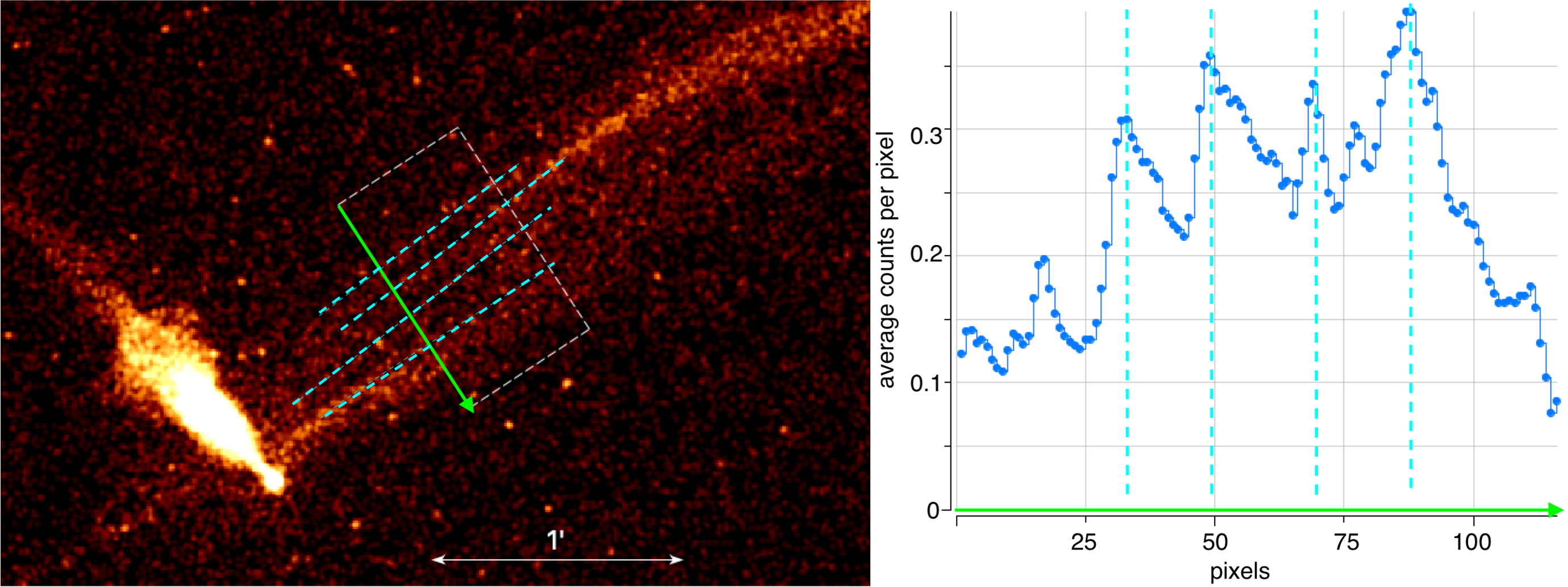}
\caption{{\sl Left:} Merged unbinned \chandra\ image (286 ks, smoothed with a $r=3$ pixel ($1\farcs5$) Gaussian kernel) showing the thread-like fine structure of the dimmer section of the misaligned outflow.  The dashed green box represents the area used to produce the brightness profile in the right panel.
{\sl Right:} Brightness profile of the portion of the outflow enclosed by the dashed white box.  The profile is taken in the transverse direction, which is shown by the green arrow.  The thin dashed cyan lines are used to highlight the apparent thread-like substructures.  One \chandra-ACIS pixel corresponds to $0\farcs492$.}
\label{fig-threads}
\end{figure*}

In Figure \ref{fig-threads} we present an un-binned \chandra\ image (cf.\ Figure \ref{fig-lighthouse-PWN}).
The image suggests that at least the dimmer portion of the main outflow, which is seen about 1$'$ northwest of the pulsar, may be composed of multiple thread-like structures which run nearly parallel to each other along the outflow (highlighted by the dashed cyan lines in Figure \ref{fig-threads}).
The threads could form as a result of filamentation (Weibel) or Bell's streaming instability (see, e.g., the X-ray ``stripes'' in the Tycho SNR forward shock; \citealt{Bykov2011}; see also Section 6.5 of \citealt{Bykov2017}). 
Alternatively, the threads may represent variations in PWN reconnection with the external (ISM) magnetic field, sampled as the pulsar travels through an inhomogeneous ISM.
The variations in reconnection can be due to variations of the external medium density, field geometry, and/or instabilities in the PWN flow. 
For example, the bubble structures seen in ${\rm H}_\alpha$ images of the Guitar PWN (PSR B2224+65; see Figure 2 of \citealt{Chatterjee2002}) indicate that the ISM density can substantially vary on scales as small as $\sim20''$ ($2.5\times10^{17}$ cm at $d=0.8$ kpc; see \citealt{Yoon2017}). 
The density variations would change the ratio of shock stand-off distance to pulsar wind particle gyroradius, thus modulating the escape rate of particles \citep{deVries2022}.
At the Lighthouse PWN's distance $d=7$ kpc, such length scales correspond to $\approx 2.5''$, which is comparable to the $\approx6''$--$8''$ separation of the thread-like structures shown in Figure \ref{fig-threads}.
It is also interesting that the threads are most prominent in the dimmer region of the outflow (about 1$'$ from the pulsar), and that the outflow returns to roughly its initial brightness shortly after (about 2$'$ from the pulsar).  
This suggests that the flow speed and/or magnetic field strength may vary substantially on parsec-scale distances along the outflow.

Another notable feature of the outflow is that it seems to originate not only from the bow shock apex but also from the tail. 
This implies that the particles are leaking out of both the tail and the bow shock apex (as appears to be the case in simulations of \citealt{Olmi2019C} -- see their Figure 5).

\subsubsection{Energetics, Flow Properties, and Magnetic Field}

The \nustar\ detection of the Lighthouse PWN marks the first detection of a misaligned outflow in the hard X-ray band (above $\sim$10 keV). 
In the 136 ks \nustar\ exposure, the outflow is clearly seen up to $\sim$25 keV, with the far reaches of it (segment 3) possibly being detected above 50 keV (though this may be unrelated background emission, as suggested by the poorer quality fit to this segment compared to the other two). 
The outflow's overall shape and size appear the same in both the \nustar\ and \chandra\ images, indicating that they do not exhibit a strong dependence on energy, at least up to $\sim$25 keV (i.e., the highest energy at which all segments are seen), which implies that the electrons emitting at higher energies do not lose most of their energy by the time they reach the farthest discernible part of the outflow (i.e., the synchrotron cooling is weak or moderate).

The strongly elongated shape of the outflow suggests that the magnetic field is predominantly oriented along the outflow, making it easier for particles to travel in that direction.
Since the shape and size of segment 3 of the outflow appear to be the same in the \chandra\ and \nustar\ images, the particle travel time along the outflow, $t_{\rm trav}$, must be smaller\footnote{We note that the following  estimate is applicable only if the particle motion across the magnetic field lines is relativistic and the gyroradii are small compared to the outflow spatial scales.  The estimate is inapplicable if the particles simply stream along the magnetic field lines.} than the synchrotron cooling time, $t_{\rm syn} \sim 1000 (E_{\rm syn}/25\ {\rm keV})^{-1/2}(B/5\ {\rm \mu G})^{-3/2}$ yrs.

From the {\sl Chandra} data, we can estimate the X-ray efficiency in the 0.5--8 keV band\footnote{We use this energy range since it allows for comparison with other misaligned outflows, as currently only the Lighthouse PWN has been studied in the hard X-ray band.}.
With the tail's luminosity $L_{\rm 0.5-8\ keV} = 7.4\times10^{33}$ erg s$^{-1}$, its efficiency is similar, $\eta_X = 5.3\times10^{-3}$.
Thus, the PWN's total X-ray efficiency $\eta_X = 1.2\times10^{-2}$. 
We note that the X-ray efficiencies of other prominent misaligned outflows, associated with PSRs B2224+65, J1509-5850, and J2030+4415, are $8\times 10^{-4}$ ($d=0.83$ kpc), $8\times 10^{-4}$ ($d=3.8$ kpc), and $2\times 10^{-4}$ ($d=0.5$ kpc), respectively, significantly lower.
The J1101 outflow's 3--79 keV efficiency is $\eta_X=1.4\times10^{-2}$.

Using the spectra and fluxes of the misaligned outflow measured in several segments, we can crudely estimate magnetic fields in those regions.  
For a PL synchrotron spectrum with photon index $\Gamma$, the magnetic field at a given magnetization parameter $\sigma= w_B/w_e$ (the ratio of magnetic to particle-kinetic energy densities) depends on the ratio of the luminosity $L(\nu_m,\nu_M)$ measured in the  $\nu_m <\nu <\nu_M$ frequency range (here, $h\nu_m=0.5$ keV and $h\nu_M = 25$), to the radiating volume $V$ (see, e.g., \citealt{Klingler2016a}):

\begin{equation}
B=\left[\frac{L(\nu_m,\nu_M)\sigma}{\mathcal{A} V} \frac{\Gamma-2}{\Gamma-1.5}\frac{\nu_1^{1.5-\Gamma}-\nu_2^{1.5-\Gamma}}{\nu_m^{2-\Gamma}-\nu_M^{2-\Gamma}} \right]^{2/7}.
\label{synch_magn_field}
\end{equation}
In this equation, $\nu_1$ and $\nu_2$  are the characteristic 
synchrotron frequencies ($\nu_{\rm syn}\simeq 3 eB\gamma^2/4\pi mc$) corresponding to the boundary energies ($\gamma_1 m_ec^2$ and $\gamma_2 m_ec^2$) of the electron spectrum ($dN_e/d\gamma \propto \gamma^{-p}\propto \gamma^{-2\Gamma+1}$; $\gamma_1<\gamma<\gamma_2$), 
and $\mathcal{A}=2^{1/2}e^{7/2}/(18 \pi^{1/2}m_e^{5/2}c^{9/2})$.  
For each misaligned outflow segment we take the average of the $\Gamma$ and normalization as measured by {\sl Chandra} and {\sl NuSTAR} (listed in Tables \ref{table-outflow-spectra} and \ref{table-outflow-chandra-spectra}), and calculate the 0.5--25 keV luminosities using those averaged values.
The \nustar\ analysis regions used in the above spectral analyses are wider than the actual width of the outflow due to its large PSF, so we use the higher-resolution \chandra\ images to estimate the volume.  
We approximate segments 1 and 2 as cylinders of radius $r=30''$ and and length $l=100''$.  
We approximate segment 3 as a sphere of radius $r=100''$.
Since $L\propto d^2$ and $V\propto d^3$, the magnetic field estimated from Equation (\ref{synch_magn_field}) weakly depends on the assumed distance, $B\propto d^{-2/7}$. 
The exact value of $\nu_2$ is not really important as long as $(\nu_1/\nu_2)^{\Gamma -1.5}\ll 1$, so we choose a plausible value $h\nu_2=25$ keV.
We assume $h\nu_1=0.5$ keV: the lowest energy at which the outflow has been observed.
It is possible that the actual $h\nu_1$ is lower, but it can not be determined with the currently available data.

For segments 1--3 we estimated $B_1 \sim 6\ \sigma^{2/7}$ $\mu$G, $B_2 \sim 6\ \sigma^{2/7}$ $\mu$G, and $B_3 \sim 4\ \sigma^{2/7}$ $\mu$G. 
The unknown lower boundary frequency $\nu_1$ is the main source of uncertainty of the magnetic field estimates for the measured spectral slopes. 
In reality, $h\nu_1$ may be lower than 0.5 keV, in which case the estimated magnetic field will be higher.
Thus, the above estimates should be considered lower limits. 
For example, if we set $h\nu_1=1$ eV, the magnetic field estimates change to $\sim$10, $\sim$14, and $\sim$10 $\sigma^{2/7}$ $\mu$G, respectively.
Also, there may be a spectral break below the lower observed frequency or lower boundary frequency, which is another reason to interpret these estimates as crude lower limits.
Lastly, we note that the magnetic field estimate for segment 3 may not be as accurate as the estimated for the other segments as segment 3 is likely not a spherical structure (as we approximated), but rather, composed of more compact filamentary structures.  
However, the existing \chandra\ data do not allow us to reliably estimate the precise morphology (and volume) of this segment of the tail; hence our spherical approximation.

One can estimate the Lorentz factors of the escaped particles as $\gamma \sim 3\times10^8 (E_{\rm syn}/8\  {\rm keV})^{1/2} (B/5\ {\rm \mu G})^{-1/2}$ corresponding to $\gamma=(1-5)\times10^{8}$ range for the 1--25 keV energies of the observed synchrotron photons.
The upper value is  a factor of 10 below the maximum $e^-/e^+$ Lorentz factor $\gamma_{\rm max}=4.8\times10^9$ ($\approx 2.4$ PeV) corresponding to the theoretical maximum accelerating potential between the pulsar's pole and light cylinder, $\Delta\Phi=(3\dot{E}/2c)^{1/2}$ \citep{Goldreich1969}. 
We note that the Guitar Nebula's pulsar, B2224+65, has much lower $\dot{E}=1.2\times 10^{33}$ erg s$^{-1}$ resulting in $\gamma_{\rm max}=1.4\times10^8$, which is somewhat below $\gamma \sim 3\times10^8 (E_{\rm syn}/8\  {\rm keV})^{1/2} (B/5\ {\rm \mu G})^{-1/2}$ for the highest energy photons observed in the misaligned outflow of the Guitar nebula. 
This lends support to the possibility that particles populating misaligned outflows may be ISM particles accelerated between the forward shock and pulsar wind termination shock ahead of the moving pulsar (see \citealt{Bykov2017}). 
In this case, the one-sidedness of the outflows may be explained by the fact that electrons and positrons would drift in opposite directions in an ordered magnetic field, and there are more electrons than positrons in the ISM ahead of the pulsar.

\begin{figure*}
\includegraphics[width=1.0\hsize,angle=0]{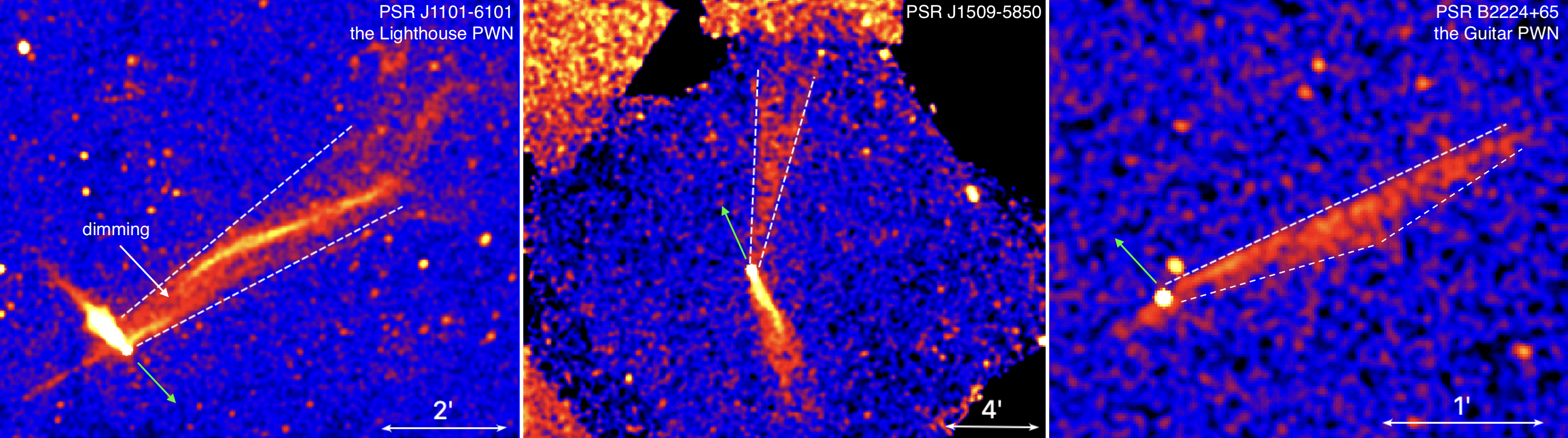}
\caption{\chandra\ images of the misaligned outflows produced by PSRs J1101--6101 (the Lighthouse PWN), J1509--5850, and B2224+65 (the Guitar PWN).  The dashed lines in the left and middle panels show the linear widening with distance seen in the Lighthouse and J1509 PWNe; the dashed line in the right panel shows the leading edge of the Guitar PWN outflow.  The green arrows mark the pulsars' directions of motion.}
\label{fig-outflows}
\end{figure*}

\subsubsection{Comparison with Other Misaligned Outflows}
While all known misaligned pulsar outflows exhibit remarkably similar spectra in the 0.5--8 keV band ($\Gamma=1.6-1.7$), they can exhibit different morphologies.
For example, the width of the Lighthouse outflow appears to increase nearly linearly with distance from the pulsar up to about $4\farcm5$ (9 pc) after which point the outflow suddenly widens, while a central (bright) part of the outflow appears to ``wiggle''.
Linear expansion with distance from the pulsar is also  clearly seen in the  PSR J1509--5850 misaligned outflow ($d_{\rm DM}\sim3.8$ kpc; $160 < v_{\perp,\rm psr} < 640$ km s$^{-1}$; see Figure \ref{fig-outflows} and \citealt{Klingler2016a}), but the rapid sideways expansion at the end is not seen (though it is not clear whether the J1509 outflow indeed lacks this behavior, or if it does expand outside the ACIS field of view), and no central (bright) interior is seen. 
Unlike the Lighthouse outflow, the J1509  and Guitar outflows do not exhibit thread-like substructures.
It is not that they can not be resolved, as the J1509 and Guitar PWNe lie at roughly half and one tenth (respectively) the distances of the Lighthouse PWN.
Unlike the J1509 and Lighthouse outflows, the Guitar  outflow ($d=0.83\pm0.14$ kpc; $v_{\perp, \rm psr} = 770\pm130$ km s$^{-1}$; \citealt{Deller2019}) first widens with distance from the pulsar but then become slightly narrower (Figure \ref{fig-outflows}).
Also, the Guitar PWN, which was so named because of its guitar-shaped H-$\alpha$ tail \citep{Chatterjee2002,Brownsberger2014,deVries2022}, lacks an X-ray tail. 
This could be due to the lower spin-down power, $\dot{E}= 1.2\times 10^{33}$ erg s$^{-1}$, and therefore a lower accelerating potential of the Guitar pulsar, such that the pulsar wind particles in the tail are not energetic enough to emit X-rays.
If this is the case, it would mean that the particle leakage mechanism (i.e., reconnection of the PWN and ISM magnetic fields) is substantially accelerating the pulsar wind, or that the particles are additionally accelerated in the transrelativistic colliding flows (between the forward shock and termination shock of the pulsar wind) ahead of the pulsar, from where they leak to the ISM \citep{Bykov2017}.

Further observations of these magnificent structures are needed to elucidate the reasons for these contrasting behaviors and to further our understanding of the complex interactions between pulsar winds and the ISM magnetic field.

\subsection{Pulsar}
The \nustar\ pulse profile of J1101 shows a single broad peak with a relatively flat top that spans about half of the phase interval, 
This is similar to the pulse profiles found with {\sl XMM-Newton} in the 0.5--10 keV band by \cite{Halpern2014} and with {\sl NICER} in the 1.5--10 keV band by \cite{Ho2022}. 
The pulsation period detected with NuSTAR is consistent with the ephemeris of \cite{Ho2022}.
We find no pulse profile dependence on energy up to at least 40 keV.

The spectrum of the pulsar$+$PWN are generally in agreement with that found by \cite{Pavan2016}. 
We also performed phase-resolved spectroscopy in two broad phase bins. 
However, due to a lack of statistics and the contaminating PWN, we do not find any statistically significant evidence of a changing photon index with phase, contrary to what has been observed in several other pulsars (e.g., \citealt{Chen2016,Hare2021}).

\section{Conclusions}
We have presented \nustar\ observations of PSR J1101--6101 and the Lighthouse PWN.
The entire outflow is clearly seen up to $\sim$25 keV, and the distal segment of it is marginally seen up to $\sim$50 keV (though the statistics are insufficient to discern whether it is unrelated background emission or synchrotron emission from the distal segment of the outflow). 
The outflow's shape and 7$'$ (14 pc) extent as seen in the 20--25 keV band (and lower energies) are consistent with those seen in the 0.5--8 keV band by \chandra\ (though the outflow's true extent may be limited by the FOV of both observatories).
We found marginal evidence of synchrotron cooling along the outflow, with the spectral slope increasing from $\Gamma\approx1.8\pm0.1$ to $\Gamma\approx2.2\pm0.1$ in the \nustar\ band.
We crudely estimated an equipartition outflow magnetic field strength for the outflow, $B \gtrsim (4-6)$ $\mu$G, which is comparable to the ISM magnetic field.

We reanalyzed archival high-resolution \chandra\ images of the Lighthouse PWN misaligned outflow to investigate its arcsecond-scale structure.  
We found evidence that at least part of the outflow is composed of multiple thread-like structures that run nearly parallel to each other, which may form from Weibel or Bell streaming instability or, more likely, from variations in PWN-ISM magnetic field reconnection.  
The \chandra\ images show that part of the outflow appears to originate from the pulsar tail rather than from the immediate vicinity of the pulsar, which may indicate that particles are leaking out of not just the bow shock apex but the pulsar tail as well.  
We also found that the outflow bends around the bow shock apex, indicating the presence of magnetic ``draping'' of the ISM magnetic field lines around the PWN bow shock.

We have performed \nustar\ timing analysis on PSR J1101--6101.
We found pulsations up to 40 keV at the frequency found by \citet{Ho2022} from recent {\sl NICER} data, and presented the pulse profile.
The consistency with the archival ephemeris suggests that this pulsar does not glitch frequently.

With \nustar\ we also detected the serendipitous X-ray source 2CXO J110158.4--605649.
We propose that it is the counterpart to the gamma-ray source 4FGL J1102.0-6054, and present evidence that supports its classification as a flat-spectrum radio quasar.

\acknowledgements
The authors wish to thank Karl Forster for assistance in planning the \nustar\ observations. 
Support for this work was provided by the National Aeronautics and Space Administration through the NuSTAR award 80NSSC21K0055. 
NK acknowledges support provided by NASA under award number 80GSFC21M0002.  
JH acknowledges support from an appointment to the NASA Postdoctoral Program at the Goddard Space Flight Center, administered by the ORAU through a contract with NASA. 
OK acknowledges support provided by NASA through Chandra Award Number G01-22055B issued by the Chandra X-ray Center, which is operated by the Smithsonian Astrophysical Observatory for and on behalf of the NASA under contract NAS8-03060.

\software{HEASoft, CIAO, Naima, Fermipy, Fermitools}

\end{document}